\documentclass[11pt,showpacs,preprintnumbers,amsmath,amssymb]{revtex4}

\usepackage{graphicx}
\usepackage{dcolumn}
\usepackage{bm}

\begin{document}

\preprint{}

\title{Twist-$3$ Distribute Amplitude of the Pion in QCD Sum Rules }
\author{Tao Huang$^{1,2}$}
\author{ Xing-Hua Wu$^2$}
 \email{xhwu@mail.ihep.ac.cn}
\author{ Ming-Zhen Zhou$^2$}
 \email{zhoumz@mail.ihep.ac.cn}
\affiliation{$^1$CCAST(World Laboratory), P.O.Box 8730,
 Beijing 100080, P.R.China}
\affiliation{$^2$Institute of High Energy Physics, P.O.Box 918
, Beijing 100039, P.R.China}
\date{\today}

\begin{abstract}
We apply the background field method to calculate the moments of
the pion two-particles twist-3 distribution amplitude (DA)
$\phi_p(\xi)$ in QCD sum rules. In this paper,we do not use the
equation of motion for the quarks inside the pion since they are
not on shell and introduce a new parameter $m_0^p$ to be
determined. We get the parameter $m_0^p\approx1.30GeV$ in this
approach. If assuming the expansion of $\phi_p(\xi)$ in the series
in Gegenbauer polynomials $C_n^{1/2}(\xi)$, one can obtain its
approximate expression which can be determined by its first few
moments.
\end{abstract}

\maketitle

\section{introduction}
The perturbative QCD theory has been applied extensively to inclusive
  and exclusive processes during the past decades. The structure
  functions in inclusive processes, $F_i(x,Q^2)$,  can be solved by matching
  the theoretical predictions with the experimental values at
  different $Q^2$ regions. Unlike the structure functions,
  the distributive amplitude in
  exclusive processes, $\phi_i(x,Q^2)$, can not easy to be
  measured directly from experiments.
  The distribution amplitudes in hadrons are the key ingredients
  and provide the universal non-perturbative inputs for many exclusive
  processes.
  Although distribution amplitudes satisfy the renormalization group equation or
  the QCD evolution equation at short distances \cite{Brodsky}, their solutions depend on
  the initial conditions $\phi_i(x,Q_0^2)$, which are determined by
  the non-perturbative theory.


  In Ref. \cite{Chernyak}, QCD sum rules were used to
  study the leading-twist distribution amplitude of the pion at the first time.
  They worked out the first few moments of the twist-2 distribution amplitude of  the pion
  and used them to restrict its shape. After that, much work
  have been done to study the leading twist distribution amplitude of
  pion and their conclusion is that the experimental data favors the
  asymptotic $\phi_{as}(\xi)=\frac{3}{2}(1-\xi^2)$ which is the solution
  of the QCD evolution equation as $Q^2 \rightarrow \infty$ .

  In the past twenty years, the studies on leading-twist distribution amplitude of
  the pion also had been done by many papers in Ref. \cite{Brodsky,Chernyak,Wang,VIZ}.
  From the naive point of view,
  the contribution of higher twist distribution amplitudes is suppressed by a factor
  $1/Q^2$ in large momentum transfer processes. In many cases,
  we can consider solely the contribution from leading-twist distribution
  amplitude, but not involve
  the impact of higher-twist distribution amplitudes. However the leading twist contribution
  to explain the data for the exclusive processes at the present energy
  region $(Q^2>4 GeV^2)$ leave a lot of space to discuss the contributions
  from higher twist terms. For example, the contribution of twist-3 distribution amplitudes
  to the pion electromagnetic form factor, although it is suppressed by
  the factor $1/Q^2$ , is comparable with and even larger
  than the contribution from leading-twist distribution amplitude of the pion at intermediate
  energy region of $Q^2$ being 2-40 $GeV^2$ in Ref \cite{Dai}.
  In the semileptonic B meson decays processes, the $B \rightarrow \pi$
  transition form factor were calculated in light cone
  QCD sum rules in the literature \cite{Braun1,Wu}.
  The $B \rightarrow \pi$ transition form factor in
  Ref \cite{Braun1} contain twist-2, twist-3 and
  twist-4 contributions.
  From the numerical results of
  Ref. \cite{Braun1} we can see that the contribution of the twist-3 distribution amplitudes
  is about
  30-50$\%$ and the contribution of the twist-4 distribution amplitudes is about
  5$\%$ to $ B \rightarrow \pi$ transition form factor.
  Although one can
  avoid to calculate the twist-3 contribution by choosing an adequate chiral
  current correlator \cite{Wu} to make the contribution of the twist-3 distribution amplitudes
  in the $B \rightarrow \pi$ transition form factor vanish automatically,
  one can't ignore their contributions once and for all.
  On the other hand, we can get the
  sum rules for the $B \rightarrow \pi$ transition form factor $f_{B\pi}^+(q^2)$
  which is determined by the twist-3 distribution amplitudes solely if choosing an
  adequate chiral correlator \cite{Zhou}. It means that the contribution from the twist-3
  distribution amplitudes to the $f_{B\pi}^+(q^2)$ has the same order of magnitude as that from the
  leading twist distribution amplitude. Therefore one has to study the twist-3
  distribution amplitudes $\phi_p(\xi),\phi_{\sigma}(\xi),\phi_{3\pi}(\alpha_i)$ and higher twist
 distribution amplitudes of the pion and other meson.

  The pion twist-3 distribution amplitudes have been studied in Ref. \cite{Braun2,Ball} in the
  chiral limit, based on the
  techniques of nonlocal product expansion and conformal expansion
  and including corrections in the meson-mass.
  Ref. \cite{Ball} studied the vector meson twist-3 distribution amplitudes in
  details.
  However they employ the equations of motion of on-shell quarks in the meson to get
  two relations among two-particles twist-3 distribution amplitudes of the pion,
   $\phi_p(\xi)$ and $\phi_\sigma(\xi)$, and three-particle
  twist-3 distribution amplitude $\phi_{3\pi}(\alpha_i)$ of the pion. Then, they
  took the moments of three-particles twist-3 distribution
  amplitude calculated in QCD
  sum rules and obtained the approximate forms for three twist-3
  distribution amplitudes of the pion. The question is
  whether one can use the equation of motion of the quark
  inside the meson since the
  quarks are not on-shell. The relations between
  $\phi_p(\xi),\phi_{\sigma}(\xi)$ and $\phi_{3\pi}(\alpha_i)$ can not be
  satisfied exactly in this point of view.

  In this paper, we do not apply the quark equation of motion and calculate the
  moments of the twist-3 distribution amplitudes $\phi_p(\xi)$
  directly in QCD sum rules approach. We employ the QCD in the
  background field \cite{ZHuang} to calculate the relevant correlator.
  The $\phi_p(\xi)$ is defined as follow:
  \begin{eqnarray}
\langle  0|\bar{d}(z)\gamma_5 ~ exp \left\{i g \int^{z}_{-z}d
x^{\mu}A_{\mu} \right \} u(-z)| \pi^+(q)\rangle =m_0^p f_\pi
\frac{1}{2}\int_{-1}^{1}d\xi \phi_p(\xi) e^{i \xi(z \cdot q)}
+\cdots  ,
\end{eqnarray}
 where the dots refer to those higher twist terms omitted here,
 and $m_0^p$ is a parameter which is to be determined in our calculation.
 This parameter has nothing to do with the masses of the quarks and/or meson.
 If we assume the expansion of $\phi_p(\xi)$ in the series in Gegenbauer polynomials
 $C^{1/2}_n(\xi)$ one can obtain the approximate
 expressions of the twist-3 distribution amplitudes of the pion which can
 be determined by their moments.

 For another twist-3 distribution amplitude $\phi_\sigma(\xi)$ , some words are in
 order. We found that, results calculated directly by QCD sum rules method depend on the
 Borel parameter seriously. That is to say, there is no way to make the $\pi$-contribution
 dominate. So we do not consider this case in this paper.

 Finally,it needs to be pointed out that, just as shown in Ref.\cite{Baulieu}, axial
 currents in a correlator would be coupled to objects like
 instantons. And this may cause
 some complication in the calculation and make the results unreliable.
 We will not explore their influences in this paper,
 some more work will be done in later paper.

 This paper is organized as follows:
 The calculation of two-point correlation function is presented
 in Sec. II.
 We calculate moments of $\phi_p(u)$ in Sec. III.
 Sec. IV is devoted to build a possible model on the twist-3 distribution amplitude.
 Sec. V is reserved for discussion and summary.

\section{calculation of two-point correlation function}

 In order to calculate two-point correlation function, we apply
 the background field approach which were expressed in Ref.
 \cite{ZHuang} explicitly. They called the QCD in background
 fields. In that framework, the non-vanishing vacuum condensates are
 described by the classical fields, while the corresponding quantum
 fields are quantized in the Furry representation and the physical
 states are built upon the physical QCD vacuum through the action
 of creation operators.
 One can derive the propagators of quarks and gluons as a
 perturbative series in gauge invariant form. The propagators are the solution
 of equations,
\begin{eqnarray*}
( i\gamma^\mu(\partial_\mu-i g A_\mu)-m )S_F(x,0)&=&\delta^4(x), \\
(g_{\mu \nu}(D_\rho D^\rho)^{AB}+2 f^{ABC}G^C_{\mu \nu})
S^{BD}_{\nu \sigma}(x,0)&=&\delta^{AD}g_{\mu \sigma}\delta^4(x).
\end{eqnarray*}
 To the aim of calculation of first few moments in this paper, we need
 only consider following terms in the two propagators respectively,
\begin{eqnarray}
S_F^{ab}(x,0) = & & \frac{i\gamma_\rho x^\rho}{2 \pi^2 x^4}\delta^{ab}+
\frac{m}{4 \pi^2 x^2} \delta^{ab}+
\frac{-\gamma^\alpha\gamma_\rho x^\rho\gamma^\beta}
     {16\pi^2 x^2}g (G_{\alpha\beta}^A(0)T^A)^{ab}
+\frac{\ln{(-x^2)}}{48 \pi^2}g \gamma^\mu
 \left[G_{\alpha\mu;}^{A\quad\alpha}(0)T^A\right]^{ab}
\nonumber \\
& &+\frac{-x^\alpha\gamma^\nu\gamma_\rho x^\rho\gamma^\mu}{48 \pi^2 x^2}g
    \left[(G_{\nu\mu;\alpha}^A+G_{\alpha\mu;\nu}^A)(0)T^A\right]^{ab},
\\
S_{\mu\nu}^{AB}(x,0) = & & \frac{g_{\mu\nu}}{4 \pi^2 x^2}\delta^{AB},
\end{eqnarray}
 where $a,b=1,2,3$,
 $(\tilde{G}_{\mu\nu})^{AB}=-if^{ABC}G_{\mu\nu}^C(A,B,C=1,2,\dots,8)$,
 $G_{\mu\nu}^A=\partial_\mu A_\nu^A-\partial_\nu
 A_\mu^A+gf^{ABC}A_\mu^BA_\nu^C$ is the classical field strength,
 $A_\mu^A$ is the classical field of the gluon.

 We define a two-point correlation function,
\begin{equation}
\Pi_{\Gamma_1 \Gamma_2}(q)=i \int d^4x e^{iq \cdot x} \langle 0
\mid T \left\{j_{\Gamma_1}(x)j_{\Gamma_2}(0) \right\} \mid 0
\rangle,
\end{equation}
 where $\Gamma_1$ and $\Gamma_1$ are Dirac matrices, and the vacuum
 state is the physical one.

 For the light quark current, the OPE of two-point correlation
 function is as follow:
\begin{eqnarray}
& & i \int d^4x e^{iq \cdot x} \langle 0 \mid T
    \left\{j_{\Gamma_1}(x)j_{\Gamma_2}(0) \right\} \mid 0 \rangle
\nonumber \\
& & =C_{\Gamma_1 \Gamma_2}^{(I)}(q^2) I +C_{\Gamma_1
    \Gamma_2}^{(m)}(q^2) \langle 0 \mid m \bar{\psi}\psi \mid 0
    \rangle +C_{\Gamma_1 \Gamma_2}^{(G^2)}(q^2) \langle 0 \mid
    \frac{\alpha_s}{\pi}G^2 \mid 0 \rangle +\cdots,
\end{eqnarray}
 where we ignore the higher power of light quark mass $m_q$ and
 only reserve the linear $m_q$ term. $C_{\Gamma_1 \Gamma_2}$'s in
 Eq.(6) are Wilson coefficients which can be worked out in perturbative QCD
 theory.

 For calculating the moments of $\phi_p(\xi)$ , we take as usual the two-point correlation
 functions:
\begin{eqnarray}
i\int d^4 x e^{iq\cdot x} \langle 0\mid T \left \{ j_5^{(2n)}(x,z)
j_5^{(0)\dag}(0) \right \} \mid 0 \rangle \equiv (z\cdot
q)^{2n}I_p^{2n,0}(q^2),
\end{eqnarray}
 where the currents are defined as
\begin{eqnarray}
& & j_5^{(2n)}(x,z)=\bar{d}(x)\gamma_5(iz\cdot \overleftrightarrow{D})^{2n} u(x),
\end{eqnarray}
with
$\overleftrightarrow{D}=\overleftrightarrow{\partial_\mu}-i2gA_\mu^a
T^a$. The relation between the light quark currents
$j_5^{(2n)}(x,z)$  and the twist-3
distribute amplitude of the pion will be given in the next section.

On the one hand, one can calculate the correlation functions in
QCD framework perturbatively. Following the familiar procedure
\cite{Wang} by applying operator expansion in the background field
approach, corresponding Feynman diagrams in Fig.1, with
condensates up to dimension-6 and the perturbative contribution part
to the lowest order, we have
\begin{eqnarray}
& & I_p^{2n,0}(-Q^2)=\frac{3}{8\pi^2}\frac{1}{2n+1}Q^2
\ln(\frac{-Q^2} {\mu^2})
-\frac{2n-1}{2}\frac{(m_u+m_d)\langle\bar{\psi}\psi\rangle}{Q^2}
\nonumber  \\
& & \quad\quad\quad\quad\quad\quad + \frac{2n+3}{24}
\frac{\langle\frac{\alpha_s}{\pi}G^2\rangle}{Q^2}-\frac{\langle\sqrt{\alpha_s}\bar{\psi}\psi\rangle^2}{Q^4}
\frac{16\pi}{81}(21+8n(n+1)) .
\end{eqnarray}

On the other hand, one can express the two-point correlation
functions in hadronic representation by inserting the complete
intermediate states with the same quantum numbers as those of
corresponding currents which is $j_5^{2n}$ in this case. By
isolating the pole term of the lowest pseudoscalar pion, the
hadronic representation can be obtained in the following,
\begin{eqnarray}
& & i\int d^4 x e^{iq\cdot x} \langle 0\mid T \left \{
j_5^{(2n)}(x,z) j_5^{(0)+}(0) \right \} \mid 0 \rangle \nonumber
\\
& & =\frac{\langle 0\mid j_5^{(2n)}(z)\mid\pi\rangle \langle \pi
\mid j_5^{(0)\dag}(0) \mid 0 \rangle} {q^2-m_\pi^2}
+\sum_{\pi^H}\frac{\langle 0\mid j_5^{(2n)}(z)\mid\pi^H\rangle
\langle \pi^H \mid j_5^{(0)+}(0) \mid 0 \rangle} {q^2-m_{\pi^H}^2}
\end{eqnarray}
where $\pi^H $s are higher exciting hadronic states.

\section{Sum rules for the moments of twist-3 distribution amplitude}

Twist-3 distribution amplitude $ \phi_{p}(\xi) $ of the pion is
given as the leading terms
by gauge invariant elements of nonlocal hadronic matrices in
Eqs.(1). Furthermore, the elements of nonlocal
hadronic matrices can be expanded near light cone $z^2=0$,
\begin{eqnarray}
\langle  0|\bar{d}(z)\gamma_5 ~ exp \left\{i g \int^{z}_{-z}d
x^{\mu}A_{\mu} \right \} u(-z)| \pi^+(q)\rangle
=\sum_{n=0}^{\infty} \frac{(-)^n}{n!} \langle 0 \mid
\bar{d}(0)\gamma_5(z \cdot \overleftrightarrow{D})^n u(0)\mid
\pi^+(q) \rangle .
\end{eqnarray}
Substituting Eq.(14) into Eq.(1), we have the following relations,
\begin{eqnarray}
\langle 0\mid j_5^{(2n)}(z)\mid\pi\rangle=f_\pi m_0^p \langle
\xi_p^{2n} \rangle (z\cdot q)^{2n}
\end{eqnarray}
where the moments of $\phi_p(\xi)$ is defined as,
\begin{eqnarray}
\langle \xi_p^{2n}
\rangle=\frac{1}{2}\int_{-1}^{1}\xi^{2n}\phi_p(\xi)d\xi .
\end{eqnarray}
Therefore the spectrum can be taken as approximately,
\begin{eqnarray}
\mathrm{Im} I_p^{2n,0}(q) =  \pi
\delta(q^2-m_\pi^2)(m^p_0)^2f^2_{\pi}\langle \xi_p^{2n} \rangle
\cdot \langle \xi^0_p \rangle+\pi
\frac{3}{8\pi^2}\frac{1}{2n+1}q^2 \theta(q^2-s_\pi),
\end{eqnarray}
where $s_{\pi}$ is the threshold parameter and should be set to a value between
$m_{\pi}^2$ and its first exciting state $\pi '$.

To get the sum rules for moments, we employ dispersion relation
for $I(Q^2)$,
$$
\frac{1}{\pi}\int ds ~ \frac{\mathrm{Im}I_p^{2n,0}(s)}{s+Q^2}
 = I_p^{2n,0}(-Q^2),
$$
and assume the contribution of the lowest state $\pi$
dominant in the sum rule. making a Borel transformation upon above
dispersion relation, we can get an improved one
\begin{equation}
\frac{1}{\pi M^2}\int ds ~ e^{-\frac{s}{M^2}}
\mathrm{Im} I_p^{2n,0}(s)=\hat{L}_M I_p^{2n,0}(-Q^2),
\end{equation}
where $M$ is the parameter used in the Borel transformation with $M=Q^2/n $ fixed,
$$
\hat{L}_{M}=\lim_{Q^2,n \rightarrow
\infty}\frac{1}{(n-1)!}(Q^2)^n(-\frac{d}{dQ^2})^n.
$$
\noindent Inserting Eq.(13) and Eq.(8) into the left and right
hand side of Eq.(14) respectively, we obtain the sum rule for
moments of $\phi_p(\xi)$,
\begin{eqnarray}
\langle \xi_p^{2n} \rangle \cdot \langle \xi^0_p \rangle
=\frac{M^4}{(m_0^p)^2 f^2_{\pi}} e^{m_{\pi}^2/M^2}
& & \left[ \frac{3}{8\pi^2}\frac{1}{2n+1} \left(
    1-(1+\frac{s_{\pi}}{M^2})e^{-\frac{s_{\pi}}{M^2}}\right) \right.
    -\frac{2n-1}{2}\frac{(m_u+m_d)\langle \bar{\psi}\psi \rangle}{M^4} \nonumber \\
& & +\frac{2n+3}{24} \frac{\langle
    \frac{ \alpha_s}{\pi}G^2 \rangle}{M^4} \left.  -\frac{16
    \pi}{81}\left( 21 + 8n(n+1) \right) \frac{\langle
    \sqrt{\alpha_s}\bar{\psi}\psi \rangle^2}{M^6} \right] .
\end{eqnarray}

As we can see, the first term in Eq.(15) is the pure perturbative
results with condensate contributions omitted completely.

\section{a possible model of the pion twist-3 distribution amplitude}

Before proceeding further we need to fix input parameters. For the
condensate parameters, we take as usual,
\begin{eqnarray}
(m_u + m_d)\langle \bar{\psi} \psi \rangle
& \simeq & -1.7 \times 10^{-4} GeV^4,
\nonumber \\
\langle \frac{ \alpha_s}{\pi} G^2 \rangle
& \simeq & 0.014 GeV^4,
\nonumber \\
\langle \sqrt{\alpha_s}\bar{\psi}\psi \rangle^2
& \simeq & 1.83 \times 10^{-4} GeV^6 .
\end{eqnarray}
The parameter $s_\pi$ in Eq.(15) should be chosen to make moments
and $m_0^p$ most stable against $M^2$ in a certain range. This can
be done by considering the ratios of $\langle
\xi^{2(n+1)}_p\rangle$ to $\langle \xi^{2n}_p\rangle$ . The
advantage is that the ratios depend only on $s_\pi$ and $M^2$, but
not $m_0^p$ and $\langle \xi^0_p \rangle$. This make us determine
$s_\pi$ and $M^2$ on a more general ground at first. The results
are shown in Fig.2. It is shown from these figures that $s_\pi$
affect the results a little and $M^2$ should be taken, at least,
$1.2 GeV^2$. The key point to avoid the concept of on-shell
equations of motion is that the introduced $m_0^p$ appearing in
the sum rule of moments. And they can be obtained through
requiring normalization of the zeroth moments($\langle \xi^0_p
\rangle =1$). With these points kept in mind, we can show the
results as ($M^2=1.5-2 GeV^2$)
\begin{eqnarray}
\begin{array}{*{3}{c@{\: \quad \:}c}}
s_\pi=1.7GeV^2 & m_0^p=1.24-1.36GeV &  \langle \xi^2_p
\rangle=0.340-0.356 &
    \langle \xi^4_p \rangle=0.167-0.210 , \\
s_\pi=1.6GeV^2 & m_0^p=1.19-1.30GeV  & \langle \xi^2_p
\rangle=0.340-0.359 &
    \langle \xi^4_p \rangle=0.164-0.211 , \\
s_\pi=1.5GeV^2 & m_0^p=1.14-1.24GeV  & \langle \xi^2_p
\rangle=0.341-0.361 &
    \langle \xi^4_p \rangle=0.160-0.212 .
\end{array}
\end{eqnarray}
The above numerical results show that the dependence of $m_0^p$
for Borel parameter is about 10\%, the dependence of $\langle
\xi^2_p \rangle$ for Borel parameter is about 5\% and the
dependence of $\langle \xi^4_p \rangle$ for Borel parameter is
about 25\% as $M^2$ is in the region we chose. It is obvious that
the fourth moment of $\phi_p(\xi)$ is more unreliable. The above
numerical results also show that the parameter $m_0^p$ is smaller
than $m_\pi^2/(m_u+m_d)\simeq 1.78-3.92GeV$ which was used in
\cite{VIZ,Braun2} when the equation of motion of on-shell quarks
are employed.

In principle, infinite number of moments are needed to determine
twist-3 distribution amplitude of the pion $\phi_p(\xi)$
completely, that is beyond one's capability by
now in this method. Also we have neglected contributions from higher
dimension condensates in the QCD calculation and higher exciting
resonances in the hadronic representation. Thus the prediction for
more moments is not reliable. We will take the above three moments into
consideration in the following analysis and expand twist-3
distribution amplitudes $\phi_p(\xi)$of the pion in Gegenbauer polynomials to
first few terms,
\begin{eqnarray}
\phi_p(\xi)=A_p + B_p ~ C^{1/2}_2(\xi)+C_p ~ C^{1/2}_4(\xi),
\end{eqnarray}
where $C^{1/2}_n(\xi)$ are Gegenbauer polynomials. $A_p,B_p,C_p $
are three coefficients to be determined through the moments given
in previous section.We obtain the two group of equations,
\begin{eqnarray}
\frac{1}{2}\int_{-1}^{1} d\xi \xi^{2n} (A_p + B_p ~
C^{1/2}_2(\xi)+C_p
~ C^{1/2}_4(\xi)) =\langle \xi_p^{2n} \rangle \quad \quad n=0,1,2.
\end{eqnarray}
Solving  the above equations, it is easy to find these
coefficients in twist-3 distribution amplitude $\phi_p(\xi)$,
\begin{eqnarray}
 \begin{array}{*{2}{l@{\quad , \quad }}l} A_p=\langle
\xi_p^0 \rangle & B_p=-\frac{5}{2}(\langle \xi_p^0 \rangle - 3
\langle \xi_p^2 \rangle) & C_p=\frac{9}{8}(3 \langle \xi_p^0
\rangle - 30 \langle \xi_p^2 \rangle + 35 \langle \xi_p^4 \rangle)
\end{array}  .
\end{eqnarray}
Substituting the value Eq.(17) into Eq.(20), the $\phi_p(\xi)$ can
be obtained.(when $(s_\pi=1.7GeV^2)$, we have, $A_p \simeq 1, B_p
\simeq (0.05)-(0.17), C_p \simeq (-1.52)-(-0.371) $. ) We can
compare Eq.(18) with the results of Ref. \cite{Braun2},
\begin{eqnarray}
\phi_p(\xi)=1 + 30 R ~ C^{1/2}_2(\xi)+\frac{3}{2}R(4\omega_{2,0}-\omega_{1,1}-2\omega_{1,0}) ~ C^{1/2}_4(\xi),
\end{eqnarray}
where $R=\frac{m_u+m_d}{m_\pi^2}\frac{f_{3 \pi}}{f_\pi}$ and the
parameters $\omega$'s can be found from the moments of $\phi_{3
\pi}$ in QCD sum rules. Comparison Eq.(21) with our results is
expressed in Fig.3. The distribution amplitude depicted in the
dashed line contains contributions from the second and the fourth
moment. But it should be pointed out that the fourth moment
calculated in this paper is less reliable than the second
one. The dotted line depicts the one only contains the
contribution from the second moment of $\phi_p(\xi)$. Finally, Our
result is closed to the asymptotic form, $\phi_p^{AS}(\xi)=1$.

\section{summary and discussion}

In applying QCD to exclusive processes, form factor and decay
amplitude are determined by the convolution of a perturbative
hard scattering amplitude with a non-perturbative distribution
amplitude of given twist. The study of the leading twist and
non-leading twist distribution amplitudes is a crucial task for
predicting exclusive processes more precisely. The asymptotic behavior
of exclusive processes is determined by the leading twist
distribution amplitude only as $Q^2$ is very large. One has to
consider the contributions from the non-leading twist at the
present energy regions. Many people have claimed that the
contributions from the twist-3 distribution amplitudes of the pion
make a sizeable corrections to the exclusive processes. Therefore
it is helpful to study the twist-3 distribution amplitudes of the
pion in different approaches in order to understand their
behaviors.

In this paper, we apply the background field approach to
 calculate  the OPE of the two-point correlation functions related
 to the moments of twist-3 distribution amplitudes of the pion.
 With the aid of the improved dispersion relations, we obtain the
 sum rules of the moments of $\phi_p(\xi)$. Inputting
 all sorts of parameters required in sum rules
 with the contribution of condensate to dimension-6, we worked out first
 three moments of $\phi_p(\xi)$ . It is shown that the main
 contribution to moments is from pertubative term,
 the contributions from condensate terms are relatively small,
 In the previous approach \cite{Braun2,Ball}, it was based on the
 techniques of nonlocal operator product expansion and conformal expansion.
 They applied the equations of motion to impose relations between
 the higher twist distribution amplitudes and $\phi_{3\pi}$ . We emphasize that the equations of
 motion can not be satisfied since the quarks are not on mass-shell
 within hadrons.

 To avoid imposing equations of motion of
 on-shell quark, it is not felicitous, we introduce a
 new parameter $m_0^p$ in the definitions of moments of $\phi_p(\xi)$ .
 Because of the emergence of the new parameter, we can not determined
 all moments if we don't know $m_0^p$ . The strategy we take
 is to require normalization condition $ \langle \xi_p^0\rangle =1 $,
 and then we obtain $m_0^p$ and other
 moments. In this paper, we have made some approximations, such as
 the lowest pole dominates and the higher dimension condensates
 are negligible, also we haven't calculated the radiative
 corrections. Based on these approximations, we obtain the parameter
 $m_0^p \simeq 1.30GeV \pm 0.06 GeV$  which is smaller than
 $m_\pi^2/(m_u+m_d)$. However, it is closed to the  phenomenological value \cite{lu}.

 Using the values of the parameter $m_0^p$ and the moments, the twist-3
 distribution amplitude  $\phi_p(\xi)$ expanded  in
 Gegenbauer polynomial can be obtained ,
 $$
 \phi_p(\xi)=1 + 0.137 ~ C^{1/2}_2(\xi)-0.721 ~ C^{1/2}_4(\xi).
 $$
 Its behavior is closed to the shapes of $\phi_p^{AS}(\xi)$ in Ref.\cite{Braun2}.

\begin{acknowledgments}
This work was supported by Natural Science Foundation of China. One
of the authors (X.H.Wu) would like to thank H.Q.Zhou for some useful
discussions.
\end{acknowledgments}

\newpage
\begin{figure}[tbp]
\begin{center}
\includegraphics*[scale=0.5,angle=0.]{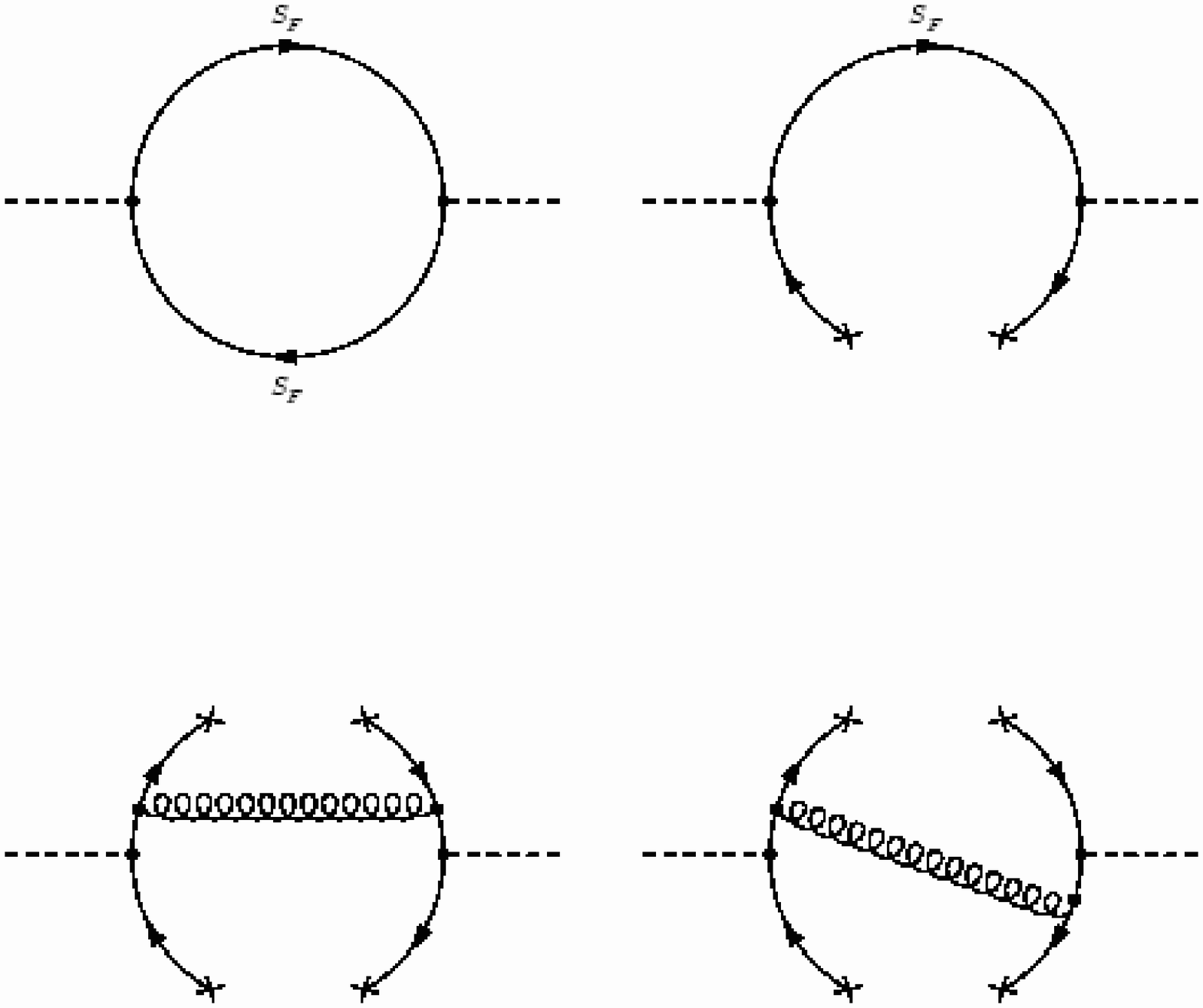}
\end{center}
\caption{Several types of Feynman diagrams calculated in this
paper. The gluon condensations are included in the Fermion
propagators implicitly and the quark condensations are depicted as
the crosses in the diagrams.}
\end{figure}

\begin{figure}[tbp]
\begin{center}
\includegraphics*[scale=1,angle=0.]{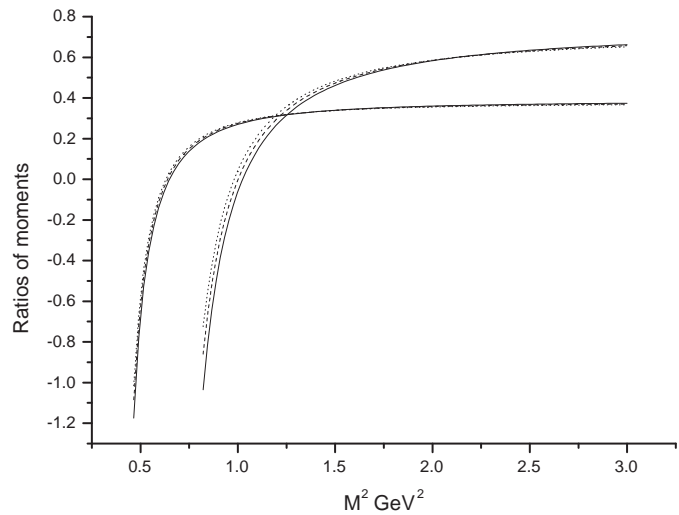}
\end{center}
\caption{The ratios of $\langle \xi^{2(n+1)}_p\rangle$ to $\langle
\xi^{2n}_p\rangle$ versus Borel parameter M. The upper curves are
the ratios with n=1 and the lower curves are the ratios with n=0,
while the solid, the dashed and the dotted curves correspond to
three threshold values: (a) $s_\pi=1.7GeV^2$; (b) $s_\pi=1.6GeV^2$
(c) $s_\pi=1.5GeV^2$, respectively. }
\end{figure}

\begin{figure}[tbp]
\begin{center}
\includegraphics*[scale=1,angle=0.]{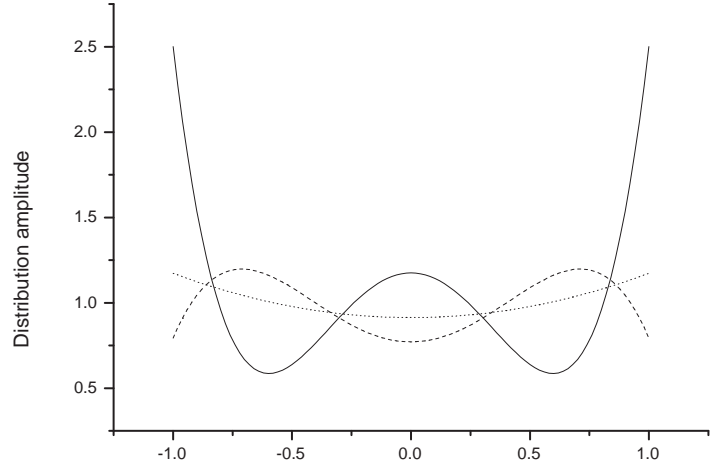}
\end{center}
\caption{The twist-3 distribution amplitude $\phi_p(\xi)$. The
solid-curve is the one of Ref.\cite{Braun2},the dashed curve
corresponds to Eq.(18) and the dotted curve corresponds to the one
only contained the second moment in this paper.}
\end{figure}

\end{document}